\documentclass[twocolumn,usenatbib,useAMS]{mn2e}

\usepackage{natbib}
\usepackage{amsbsy}
\usepackage{amssymb}
\usepackage{amsmath}
\usepackage{graphicx}
\usepackage[caption = false]{subfig}
\usepackage [english]{babel}
\usepackage [autostyle, english = american]{csquotes}
\usepackage{natbib}
\MakeOuterQuote{"}

\begin{document}

\title[Minimum Glitch and Crustquake]{Minimum Glitch of the Crab Pulsar and the Crustquake as a Trigger Mechanism}

\author[Akbal, Alpar]{ O. Akbal$^1$ \& M.A. Alpar$^{1}$ \\
  \\
  $^1$Sabanc{\i} University, Faculty of Engineering and Natural Sciences, Orhanl{\i}, 34956 Istanbul, Turkey}

\maketitle

\begin{abstract}
We discuss the minimum glitch size of Crab observed by \cite{espinoza14}. Modelling the crustquake as a trigger mechanism we estimate the size of the broken plate. The plate size obtained, D $\sim 100 m$ is comparable to plate size estimates for PSR J1119--6127. The plate size naturally leads to an estimate of the number of unpinned vortices involved in the glitch, $N \sim 10^{13} $. This number is of the same order in all Crab and Vela pulsar glitches. The mimimum glitch relates the constancy of all these numbers among different pulsars to the basic plate size involved in crust breaking. This in turn depends on the critical strain angle $\theta_{cr}$ of the Coulomb lattice in the neutron star crust. The minimum glitch size implies $\theta_{cr} \sim 10^{-1}$ in agreement with theoretical and computational estimates.  
\end{abstract}


\section{Introduction}
\label{sec:intro}
Pulsars have highly stable rotation rates. They also commonly show timing irregularities in two ways: (i) continuous stochastic deviations from the simple slowdown model ("timing noise"), (ii) abrupt changes in their rotation rates and spin-down rates ("glitches"). Glitch sizes vary by some orders of magnitude ($10^{-12}< \Delta \Omega / \Omega < 10^{-5}$) \citep{espinoza11} with power law distributions \citep{melatos08}. Several theoretical models have been proposed to explain the large glitches (see \cite{haskell15} for a review). Even if triggered by crust breaking, glitches are amplified to large magnitudes by the sudden unpinning of vortex lines in the neutron star crust superfluid \citep{anderson75}. \cite{alpar93} proposed an explanation of the maximum glitch size by the involvement of all vortices that can unpin. As for the minimum glitch size, it is hard to resolve the smallest glitches from the "timing noise" events at the lower end of the distributions. Until recently it was not known whether a minimum glitch size existed. 

\cite{espinoza14} attained a level of detection sensitivity that exposed the smallest events of the Crab pulsar which are distinct from timing noise and can be identified as glitches. Before this work the smallest glitches could not be distinguished from the timing noise. \cite{espinoza14} built an "automated glitch detector" to uncover the full distribution of glitch sizes of the Crab pulsar and determined the smallest glitch size as $ \Delta \Omega / \Omega \sim 1.7 \times 10^{-9}$, by distinguishing a resolved glitch from timing noise, as an abrupt positive step in rotational frequency ($ \Delta \Omega > 0 $) together with a discrete negative (or null) step in spin-down rate. Using X-ray data, \cite{vivekanand16} also reported that Crab might have exhibited another, slightly smaller glitch event ($ \Delta \Omega / \Omega \sim 1.3 \times 10^{-9} $). 

Glitches are thought to invoke an abrupt angular momentum transfer from the pinned superfluid component which rotates slightly faster than the solid crust \citep{anderson75}. The quantized vortices in the superfluid component interact with the nuclei in the lattice over distinct pinning regions. A lag develops between superfluid and crust angular velocities, and finally reaches a critical threshold at which the vortices are abruptly unpinned in an avalanche and promptly move outward in the direction away from the rotation axis, leading to speeding up of the crust (the glitch).

A crustquake might play a significant role as a trigger mechanism for vortex unpinning. Spinning down of the star \citep{ruderman69, baym71}, internal electromagnetic strains \citep{lander14}, or the vortex-lattice interaction \citep{ruderman76, chamel06, chau93} will cause stresses on the crust. The crust has a maximum strain beyond which it can no longer sustain elastic deformations. This leads to seismic activities, i.e. quakes. In the spinning down of the star, the fluid core can modify its shape from an oblate spheroid to more spherical form, while the solid crust is strained and has to break to readjust its shape as the strain angle reaches a critical value. The crust is expected the break along fault lines, resulting in motion of broken plates. Locally, the plates undergo plastic distortion until fracture when the strain reaches a critical value, the critical strain angle, $ \theta_{cr} $ \citep{jones03}. The fractional change in the moment of inertia of the crust is related to the fractional changes in the rotation rate $ \Omega $ and the spin-down rate $ \dot{\Omega} $:

\begin{equation}
\frac{\Delta \Omega}{\Omega}=-\frac{\Delta I}{I}=\frac{\Delta \dot{\Omega}}{\dot{\Omega}}.
\end{equation}
For a pure crustquake (without any associated vortex motion) the first equality reflects the conservation of angular momentum. The decrease of the inertial moment produces an increase in the angular velocity. The second equality reflects the constancy of the external torque. This is the case for all Crab pulsar glitches, and indeed for all pulsar glitches except for a few observed from PSR J1119--6127 \citep{weltevrede11}, PSR J1846--0258 \citep{livingstone10} and RRAT J1819--1458 \citep{lyne09}. 

There are no Crab pulsar glitches that can be modelled as pure crustquakes: Observations show that $ \Delta \dot{\Omega}/\dot{\Omega} $ is many orders of magnitude larger than $ \Delta \Omega / \Omega $. Since $ \Delta \Omega / \Omega \sim 10^{-9} \ll \Delta \dot{\Omega} / \dot{\Omega} \sim 10^{-4} $, the minimum glitch as well as other Crab pulsar glitches should not be only due to a crustquake. So there must be another agent involved in the changes in $ \Omega $ and $ \dot{\Omega} $. The sudden unpinning of pinned vortices in neutron star superfluids \citep{anderson75} can explain the observed values of $ \Delta \Omega / \Omega $ and $ \Delta \dot{\Omega}/\dot{\Omega} $ \citep{alpar93, alpar96}. Starquakes can act as triggers of unpinning. As these unpinned vortices can induce further vortex unpinning, larger glitches are produced as avalanches. This scenario is expected in the "middle aged" (older than $ 10^{4} $ years), "Vela-like" pulsars in which the network of vortex pinning regions is sufficiently connected. The "Crab-like" pulsars with young age cannot produce larger glitch events due to lack of connections between vortex traps (potential unpinning sites). The model of glitch induced crustquakes is especially favourable for the glitches of Crab. The "persistent shifts" in the spin-down rate commonly observed in the Crab pulsar's glitches are interpreted as evidence for the formation of new vortex traps \citep{alpar96}. Such persistent shifts are probably occurring also in the Vela and other pulsars \citep{akbal17}. Various pinning forces produce inhomogeneities in the vortex density distribution in the crustal lattice. Vortex traps are built at locations of particularly strong pinning forces, inducing an excess density of pinned vortices. Excess vortex density sets up local superfluid currents circulating around the traps. This makes pinning difficult in areas surrounding the vortex traps, setting up vortex depletion regions between adjacent traps. Pinning forces are extra strong within the trap but not around it. Within the traps, where vortex creep continues, as the critical conditions for unpinning are reached, the vortices collectively unpin through the vortex free regions and induce the discharge of vortices from other traps. The vortex depletion regions contribute to angular momentum transfer at glitches, while they do not contribute to spin-down between glitches since they do not support a vortex density and vortex creep.

The vortex creep model finds a natural analogy in electric circuits. The angular velocity lag $\omega$ drives the vortex current, analogous to the voltage in electric circuit. Vortex depletion regions (with moment of inertia $I_{B}$ in Equation (10) below) do not sustain any vortex currents; here vortex motion can take place only during glitches, just like an electrical capacitor which does not transmit electric current except in discharges. The vortex creep regions (with moment of inertia $I_{A}$ in Equation (10) below) behave like resistive circuit elements which connect capacitive regions and contribute to both angular momentum transfer at the time of glitch and to the continues spin down rate between glitches due to creep.

New vortex traps with surrounding vortex free regions can be created in a crustquake. The newly defined vortex depletion regions were contributing to vortex creep and therefore to spin down before the glitch, while they no longer sustain creep and contribute to spin down after the glitch. Hence there will be a step in the spin-down rate, $ \Delta \dot{\Omega}_{p} $, the persistent shift, which is not going to heal.

The glitch is likely to involve sudden unpinning of vortices in addition to the triggering crustquake. The glitch magnitudes $ \Delta \Omega / \Omega $ and $ \Delta \dot{\Omega}/\dot{\Omega} $ are both affected in different ways, by the inertial moment of the superfluid regions through which the vortices move at the glitch, as well as the quake associated physical change in the moment of inertia, so that

\begin{equation}
\frac{\Delta \Omega}{\Omega}_{(min)} \geq \frac{\Delta I}{I}_{(quake)}.
\end{equation}

\section{Geometry of the Crustquake and Some Estimates}

If glitches are triggered by crustquakes, the minimum glitch size observed can be used to obtain information on crust properties. Modeling the geometry of crust breaking, we relate the change in the moment of inertia with the crustal plate's size, the critical strain angle, and the number of vortices carried by the plate in larger events.





\begin{figure}
\centering
\vspace{0.5cm}
\includegraphics[width=1.0\linewidth]{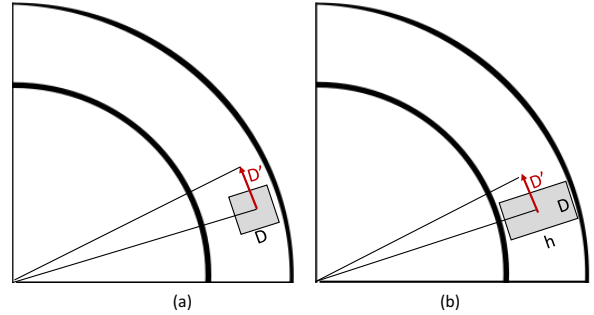}
\caption{Cartoons (cross section) of crust breaking in spin-down of the neutron star: a) a cubic "plate" of dimensions $ D \times D \times D $ moves by a distance $D' < D$ towards the rotation axis; b) a prism, $ D \times D \times h $ moves by a distance $D' < D$ towards the rotation axis.}
\label{fig:geometry}
\end{figure}

Some vortices perturbed by the quake get unpinned. If these vortices move downstream in the azimuthal direction and induce other vortices to unpin in the slightly outer regions, larger glitches can be created by the amplification of vortex motion. The minimum glitch magnitude ($ \Delta \Omega / \Omega = 1.7 \times 10^{-9}) $ and the magnitude of change in spin-down $ (\Delta \dot{\Omega} / \dot{\Omega}=1.2 \times 10^{-4}) $ imply that this event involves both a crustal quake and a vortex avalanche. While the breaking and movement of crust are rigid and irreversible processes, the superfluid component's spin-down by vortex flow (creep) can relax back to the pre-glitch states in time.

\subsection{Size of the Broken Plate and the Critical Strain}
The size of the broken plate can be found by relating it with the change in the moment of inertia of the region where the crustquake takes place. We study this in two geometries. 
 
In the first model a cubic crustal plate with volume $ V=D^{3} $ moves towards higher latitude by a distance of $ D' $ (Figure 1a). This movement results in a fractional decrease $ \Delta I/I $ in the moment of inertia of the solid (taken to be incompressible):

\begin{equation}
\dfrac{\Delta I}{I}=\dfrac{m[R^{2}-(R-D')^{2}]}{\frac{2}{5}MR^{2}}
\end{equation} 
where $ R $ is the radius, $ M $ is the mass of the entire star, $ m $ is the mass of the moving plate, and $ D' $ is the distance along which the plate flits during the crustquake. $ m $ can be related with the volume of plate as $ m=\rho D^{3} $ where $ D $ is the broken plate size and $ \rho $ is the mass density. $ D $ and $ D' $ are related by a factor $ f $, ($ D'=fD $), the ratio between the plate size and the distance the plate moves. The distance that the broken plate moves during the crustquake, $ D' $, corresponds to a change of the inclination angle, $ \Delta \alpha $, between the rotation and magnetic axes as the crustquake takes place. This change could be likely observed in some glitches, as a change in the external electromagnetic torque during an extraordinary glitch event. In the PSR 1119-6127 \citep{antonopoulou15, akbal15}, with large magnetic field ($ \sim 10^{13} $ G), which produce the crustquake at a location very close to the magnetic axis and/or to the surface of the star, the crustal movement amplified by the elastic response of the field lines can easily affect the magnetospheric activities. Such a glitch associated change in the external torque is not observed in Crab, so we cannot determine the shift in the inclination angle, $ \Delta \alpha $, and $ D' $ in this way. Scaling $ \Delta \alpha \sim 10^{-4} $ from the result of \cite{akbal15} for PSR 1119-6127, and using the relation $ D'=R \Delta \alpha=fD $ with the estimates of $ D $ we find

\begin{equation}
f \sim 10^{-2} \left( \frac{R_{6} \Delta \alpha_{-4}}{D_{4}} \right).
\end{equation}   
The broken plate moves towards the rotational axis, with a distance smaller than its size, by a factor of $ \sim 10^{-2} $, during the crustquake.

Using (2) and (3) with these assumptions and the minimum glitch size, $ \Delta \Omega / \Omega=1.7 \times 10^{-9} $ measured by \cite{espinoza14} we find

\begin{equation}
1.7 \times 10^{-9} \gtrsim \frac{5 \rho fD^{4}}{MR},
\end{equation}

\begin{equation}
D \lesssim 1.6 \times 10^{4}\left( \frac{MR_{6}}{fM_{\odot}\rho_{13}} \right)^{1/4} cm,  
\end{equation}
where $ \rho_{13} $ is the crustal mass density in units of $ 10^{13} $ g cm$ ^{-3} $, $M_{\odot} = 2 \times 10^{33}$ g is the solar mass, and $ R_{6} $ is the neutron star radius in units of $ 10^{6} $ cm.

Next we consider a plate of volume $ V=D^{2}h $ where $ h $ is the thickness (depth) of the broken plate (Figure 1b). The size of broken plate is related to the critical strain angle by 

\begin{equation}
D \sim \theta_{cr} h.
\end{equation}
We obtain the bound

\begin{equation}
D \lesssim 7.6 \times 10^{3} \left( \frac{MR_{6} \theta_{cr,-1}}{fM_{\odot}\rho_{13}} \right)^{1/4} cm. 
\end{equation}
using the minimum glitch size. The crustquake occurs at the critical strain angle, $ \theta_{cr} $, at which point the yielding takes place in the crustal lattice. On theoretical grounds the critical strain angle is estimated to be $ \theta_{cr} \sim 10^{-2}-10^{-1} $ (Akbal \& Alpar in preparation). This is in agreement with the results of the molecular dynamical simulations by \cite{HK2009} and \cite{hoffman12}.

A third model can be based on the idea that the crust might fail collectively as a result of large strain. This model involves many broken plates of size $D$ within a crustal ring, which is close to the equatorial region, of radius R, with each plate moving a distance $D' = fD$, so that the volume is taken to be $V = 2πRDh$. This gives a plate size $ D\lesssim 10.5 $ meters.

The flares and bursts, sometimes associated with the glitches in magnetars, are also thought to be the manifestations of crustquakes \citep{thompson95}. The magnetic energy for a typical radio pulsar cannot reach the elastic energy of the crust, while for a highly magnetized neutron star with the dipole magnetic field at least $ \sim 10^{15} $ G, the Lorentz force is comparable to the elastic force, hence it is quite likely that the magnetic stresses could provoke seismic activity. The geometry of poloidal and toroidal components determine where the crust breaking takes place. \cite{lander14} found that a strong toroidal component is responsible for the yielding, and the characteristic local magnetic field strength should be $ \sim 10^{14}-10^{15} $ G so that even the most luminous magnetar giant flares can be explained by the crustal energy release of magnetic stresses. Explaining the short burst events with energies up to $ 10^{41} $ erg in magnetars in terms of the crustal yielding, they also found the depth of fracture in magnetars crust as $ \sim 20 $ m, comparable with our results for the plate size.

\subsection{Number of Vortices Involved in a Larger Glitch}

Vortices unpinned from a broken crustal plate can induce other vortices to unpin if they encounter more pinned vortices close to critical conditions for unpinning. In older pulsars which exhibit larger glitches connected vortex trap regions are already established in the sense that each batch of already unpinned vortices, before they equilibrate with the background superfluid flow, are likely to encounter other batches of vortices ('vortex traps') which are close to critical conditions for unpinning. Each unpinned batch perturbs other vortices so that the avalanche continues like a relay race. This vortex avalanche does not build up efficiently in the younger pulsars exhibiting smaller glitches, because their vortex traps are not yet numerous and connected \citep{alpar93}.

Assuming that some vortices carried by the broken plate are unpinned, we can also estimate the number of unpinned vortices, $ N $, triggered by the crustquake, using the relation

\begin{equation}
N=D^{2} \left( \frac{2\Omega}{\kappa} \right).
\end{equation}    
Here $ D^{2} $ is the area of the plate, $ \Omega $ is the angular velocity of star, and $ \kappa $ is the quantum vorticity. $ 2 \Omega / \kappa $ gives the area density of vortices. These give the estimates for the number of vortices depending on the fracture geometry as: $ N_{cube} \lesssim 5 \times 10^{13} $, $ N_{prism} \lesssim 1.1 \times \times 10^{13} $.

Angular momentum conservation states that \citep{alpar93}:

\begin{equation}
\frac{\Delta \Omega}{\Omega}=\left( \frac{I_{A}}{2I}+\frac{I_{B}}{I} \right) \frac{\delta \Omega}{\Omega}
\end{equation} 
where $ I_{A}/I $ and $ I_{B}/I $ are the fractional moments of inertia of superfluid regions through which the unpinned vortices move rapidly during the glitch. Here $ I_{A}/I $ refers to the vortex creep regions and $ I_{B}/I $ to the vortex depletion regions. $ \delta \Omega $ is the change in the superfluid rotation rate due to this sudden motion of $ N $ vortices:

\begin{equation}
N=\frac{2\pi R^{2} \delta \Omega}{\kappa}.
\end{equation}
Separate analysis of angular momentum balance in both Vela \citep{alpar93} and Crab \citep{alpar96} pulsars' glitches, and also in PSR J1119-6127 \citep{akbal15}) typically yields the same $ N \sim 10^{13} $. It is significant that the $ N $ estimates we obtain here; as the number of vortices participating in the avalanche with crust breaking indicated by the minimum glitch is in agreement with these earlier results. The reason of this common number in pulsar glitches, regardless of the pulsar age, the glitch size or the size of pinning and creep regions, has not been well known. Now we propose that this common scale of $ N $ is related to the broken plate size $ D $, is therefore determined, like $ D $, by the critical strain angle $ \theta_{cr} \sim 10^{-1} $.

\subsection{A much smaller glitch from a millisecond pulsars}
Recently a glitch that is two orders of magnitude smaller ($ \Delta \Omega / \Omega = 2.5 \times 10^{-12}$) has been observed in the millisecond pulsar, PSR J0613-0200 by \cite{mckee16}. The crustquake model we propose here gives the estimates of its broken plate size as $ D_{cube} \cong 38 $ m, $ D_{prism} \cong 13.6 $ m for two different geometries. The critical strain angle and the number of vortices are $ \theta_{cr} \sim 10^{-2} $ and $ N \sim 10^{11}-10^{12}$. We suggest that these different estimates for a millisecond pulsar are due to its very old age. A millisecond pulsar will have experienced many glitches and crustquakes, as well as accreting for a long time so that their crusts may have been distorted many times. Hence the critical strain angle has been reduced (annealed) to maybe $\theta_{cr} \sim 10^{-2}$, reducing the plate size $ D $ and the number of unpinned vortices proportionately to give $ N \sim 10^{11}-10^{12}$. The impact of the impurities and defects in an accreted NS crust was already investigated by \cite{HK2009, hoffman12}, who concluded that the critical strain is reduced substantially with the presence of the defects and impurities in the crustal lattice as a result of accretion.


\section{Conclusions}
\label{sec:conclusions}
The minimum glitch size from the Crab pulsar, resolved from the timing noise \cite{espinoza14} gives the opportunity to evaluate this event in terms of the crustquake as a glitch trigger. The change in the inertial moment is directly related with this minimum glitch size in the pure crustquake model. Introducing some breaking geometries we find an upper limit for the size of the broken plate (Eqs. (6) and (8)), with some scaling factors like the star's mass, radius, critical strain angle. The plate size is then used to obtain the number of vortices, $ N $, taking part in glitches. These estimates for the Crab pulsar are $ D_{cube} \sim 160 $ m, $ D_{prism} \sim 75 $ m, depending on geometrical assumptions, for the plate size, and give the order of $ N \sim 10^{13}$ for the number of vortices that unpin in a glitch. The plate sizes we estimate here are roughly in agreement with the result of \cite{akbal15} for PSR J1119--6127. Our estimate for the plate sizes are also comparable to the height of the ``mountain'', estimated for the Crab pulsar \citep{chamel08}

The number $ N \sim 10^{13} $ is typically obtained for glitches of all different sizes, in particular for the Crab and Vela pulsar glitches, in the framework of the vortex creep-unpinning models. The agreement of $ N $ associated with one initial plate that triggers the glitch explains why all glitches have the same number of vortices. The qualitative difference of the glitch behaviour of Vela and Crab is due to evolutionary reasons. The moment of inertia involved in a glitch increases with age \citep{pines85}. While Vela and older pulsars should already have a well connected creep network so that the large glitches can be created, Crab is still in the stage of creation of the creep regions. While the common number of vortices taking part in glitch is explained by the size of the initial broken plate, the glitch magnitude depends also on the moment of inertia of the superfluid regions through which the unpinned vortices move in a relay.

Glitches are likely triggered by crust breaking. The critical strain angle, $ \theta_{cr} $, is the significant quantity which determines the size of the broken plate. The minimum glitch size is related with the inertial moment of the region, in which the crustquake takes place, and the size of the broken plate. The broken crust plate size in turn determines the number of vortices involved in the unpinning avalanche that effects the size of the amplified glitch. This number, $N \sim 10^{13} $, is typical of all small and large glitches from Vela, Crab and PSR J1119--6127 analysed so far in terms of vortex unpinning, pointing out a particular scale of the glitch trigger. The coincidence of the number of vortices is highly suggestive and is explained in terms of crust breaking with a typical plate size and critical strain angle which is found to be in agreement with theoretical and computational estimates $ \theta_{cr} \sim 0.1 $, corresponding to the unscreened Coulomb lattice in the neutron star crust. We conclude that, fundamentally, the minimum glitch size and the regularity in the number of vortices involved in glitches are due to the large critical strain angle, $ \theta_{cr} \sim 0.1 $, of the neutron star crust.

\section*{Acknowledgments}
This work is supported by the Scientific and Technological Research Council of Turkey
(T\"{U}B\.{I}TAK) under the grant 117F007. M. Ali Alpar is a member of the Science Academy (Bilim Akademisi), Turkey. We thank the referee for useful comments.

\bibliographystyle{mn2e}
\bibliography{crab_ref}

\providecommand{\noopsort}[1]{}\providecommand{\singleletter}[1]{#1}%
\begin{thebibliography}{27}
\expandafter\ifx\csname natexlab\endcsname\relax\def\natexlab#1{#1}\fi

\bibitem[{{Akbal} {et~al}\mbox{.}(2017){Akbal}, {Alpar}, {Buchner}, \&
  {Pines}}]{akbal17}
{Akbal} O., {Alpar} M.~A., {Buchner} S., {Pines} D., 2017, mnras, 469, 4183

\bibitem[{{Akbal} {et~al}\mbox{.}(2015){Akbal}, {G{\"u}gercino{\u g}lu}, {{\c
  S}a{\c s}maz Mu{\c s}}, \& {Alpar}}]{akbal15}
{Akbal} O., {G{\"u}gercino{\u g}lu} E., {{\c S}a{\c s}maz Mu{\c s}} S., {Alpar}
  M.~A., 2015, mnras, 449, 933

\bibitem[{{Alpar} {et~al}\mbox{.}(1993){Alpar}, {Chau}, {Cheng}, \&
  {Pines}}]{alpar93}
{Alpar} M.~A., {Chau} H.~F., {Cheng} K.~S., {Pines} D., 1993, apj, 409, 345

\bibitem[{{Alpar} {et~al}\mbox{.}(1996){Alpar}, {Chau}, {Cheng}, \&
  {Pines}}]{alpar96}
{Alpar} M.~A., {Chau} H.~F., {Cheng} K.~S., {Pines} D., 1996, apj, 459, 706

\bibitem[{{Anderson} \& {Itoh}(1975)}]{anderson75}
{Anderson} P.~W., {Itoh} N., 1975, nat, 256, 25

\bibitem[{{Antonopoulou} {et~al}\mbox{.}(2015){Antonopoulou}, {Weltevrede},
  {Espinoza}, {Watts}, {Johnston}, {Shannon}, \& {Kerr}}]{antonopoulou15}
{Antonopoulou} D., {Weltevrede} P., {Espinoza} C.~M., {Watts} A.~L., {Johnston}
  S., {Shannon} R.~M., {Kerr} M., 2015, mnras, 447, 3924

\bibitem[{{Baym} \& {Pines}(1971)}]{baym71}
{Baym} G., {Pines} D., 1971, Annals of Physics, 66, 816

\bibitem[{{Chamel} \& {Carter}(2006)}]{chamel06}
{Chamel} N., {Carter} B., 2006, mnras, 368, 796

\bibitem[{{Chamel} \& {Haensel}(2008)}]{chamel08}
{Chamel} N., {Haensel} P., 2008, Living Reviews in Relativity, 11, 10

\bibitem[{{Chau} \& {Cheng}(1993)}]{chau93}
{Chau} H.~F., {Cheng} K.~S., 1993, prb, 47, 2707

\bibitem[{{Espinoza} {et~al}\mbox{.}(2014){Espinoza}, {Antonopoulou},
  {Stappers}, {Watts}, \& {Lyne}}]{espinoza14}
{Espinoza} C.~M., {Antonopoulou} D., {Stappers} B.~W., {Watts} A., {Lyne}
  A.~G., 2014, mnras, 440, 2755

\bibitem[{{Espinoza} {et~al}\mbox{.}(2011){Espinoza}, {Lyne}, {Stappers}, \&
  {Kramer}}]{espinoza11}
{Espinoza} C.~M., {Lyne} A.~G., {Stappers} B.~W., {Kramer} M., 2011, mnras,
  414, 1679

\bibitem[{{Haskell} \& {Melatos}(2015)}]{haskell15}
{Haskell} B., {Melatos} A., 2015, International Journal of Modern Physics D,
  24, 1530008

\bibitem[{{Hoffman} \& {Heyl}(2012)}]{hoffman12}
{Hoffman} K., {Heyl} J., 2012, mnras, 426, 2404

\bibitem[{{Horowitz} \& {Kadau}(2009)}]{HK2009}
{Horowitz} C.~J., {Kadau} K., 2009, Physical Review Letters, 102, 191102

\bibitem[{{Jones}(2003)}]{jones03}
{Jones} P.~B., 2003, apj, 595, 342

\bibitem[{{Lander} {et~al}\mbox{.}(2015){Lander}, {Andersson}, {Antonopoulou},
  \& {Watts}}]{lander14}
{Lander} S.~K., {Andersson} N., {Antonopoulou} D., {Watts} A.~L., 2015, mnras,
  449, 2047

\bibitem[{{Livingstone}, {Kaspi} \& {Gavriil}(2010){Livingstone}, {Kaspi}, \&
  {Gavriil}}]{livingstone10}
{Livingstone} M.~A., {Kaspi} V.~M., {Gavriil} F.~P., 2010, apj, 710, 1710

\bibitem[{{Lyne} {et~al}\mbox{.}(2009){Lyne}, {McLaughlin}, {Keane}, {Kramer},
  {Espinoza}, {Stappers}, {Palliyaguru}, \& {Miller}}]{lyne09}
{Lyne} A.~G., {McLaughlin} M.~A., {Keane} E.~F., {Kramer} M., {Espinoza} C.~M.,
  {Stappers} B.~W., {Palliyaguru} N.~T., {Miller} J., 2009, mnras, 400, 1439

\bibitem[{{McKee} {et~al}\mbox{.}(2016){McKee}, {Janssen}, {Stappers}, {Lyne},
  {Caballero}, {Lentati}, {Desvignes}, {Jessner}, {Jordan}, {Karuppusamy},
  {Kramer}, {Cognard}, {Champion}, {Graikou}, {Lazarus}, {Os{\l}owski},
  {Perrodin}, {Shaifullah}, {Tiburzi}, \& {Verbiest}}]{mckee16}
{McKee} J.~W. {et~al.}, 2016, mnras, 461, 2809

\bibitem[{{Melatos}, {Peralta} \& {Wyithe}(2008){Melatos}, {Peralta}, \&
  {Wyithe}}]{melatos08}
{Melatos} A., {Peralta} C., {Wyithe} J.~S.~B., 2008, apj, 672, 1103

\bibitem[{{Pines} \& {Alpar}(1985)}]{pines85}
{Pines} D., {Alpar} M.~A., 1985, nat, 316, 27

\bibitem[{{Ruderman}(1969)}]{ruderman69}
{Ruderman} M., 1969, nature, 223, 597

\bibitem[{{Ruderman}(1976)}]{ruderman76}
{Ruderman} M., 1976, apj, 203, 213

\bibitem[{{Thompson} \& {Duncan}(1995)}]{thompson95}
{Thompson} C., {Duncan} R.~C., 1995, mnras, 275, 255

\bibitem[{{Vivekanand}(2016)}]{vivekanand16}
{Vivekanand} M., 2016, aap, 586, A53

\bibitem[{{Weltevrede}, {Johnston} \& {Espinoza}(2011){Weltevrede}, {Johnston},
  \& {Espinoza}}]{weltevrede11}
{Weltevrede} P., {Johnston} S., {Espinoza} C.~M., 2011, mnras, 411, 1917

\end{thebibliography}
\addcontentsline{toc}{chapter}{\textsc{bibliography}}
\end{document}